\documentclass[aps,pra,twocolumn,showpacs]{revtex4-1} 
\usepackage[latin9]{inputenc}
\usepackage{color}
\usepackage[colorlinks,urlcolor=blue,citecolor=blue,linkcolor=blue]{hyperref}
\usepackage{verbatim}
\usepackage{float}
\usepackage{amsmath}
\usepackage{amssymb}
\usepackage{graphicx}
\usepackage{esint}
\usepackage[normalem]{ulem} 
\usepackage{natbib}
\begin{document}

\title{Lifetime of Feshbach dimers in a Fermi-Fermi mixture of $^6$Li and $^{40}$K}

\author{M. Jag$^{1,2}$}
\author{M. Cetina$^{1,2}$}
\altaffiliation[Present address:]
{ Joint Quantum Institute, University of Maryland Department of Physics and National Institute of Standards and Technology, College Park, MD, 20742, USA}
\author{R. S. Lous$^{1,2}$}
\author{R. Grimm$^{1,2}$}
\affiliation{$^1$Institut f\"ur Quantenoptik und Quanteninformation (IQOQI), \"Osterreichische Akademie der Wissenschaften, 6020 Innsbruck, Austria}
\affiliation{$^2$Institut f\"ur Experimentalphysik, Universit\"at Innsbruck, 6020 Innsbruck, Austria}

\author{J. Levinsen$^{3}$}
\author{D. S. Petrov$^{4}$}
\affiliation{$^3$School of Physics and Astronomy, Monash University, Victoria 3800, Australia}
\affiliation{$^4$LPTMS, CNRS, Univ.\ Paris Sud, Universit\'e Paris-Saclay, 91405 Orsay, France}

\date{\today}

\begin{abstract}
We present a joint experimental and theoretical investigation of the lifetime of weakly bound dimers formed near narrow interspecies Feshbach resonances in mass-imbalanced Fermi-Fermi systems, considering the specific example of a mixture of $^6$Li and $^{40}$K atoms. 
Our work addresses the central question of the increase in the stability of the dimers resulting from Pauli suppression of collisional losses, which is a well-known effect in mass-balanced fermionic systems near broad resonances. 
We present measurements of the spontaneous dissociation of dimers in dilute samples, and of the collisional losses in dense samples arising from both dimer-dimer processes and from atom-dimer processes.
We find that all loss processes are suppressed close to the Feshbach resonance. 
Our general theoretical approach for fermionic mixtures near narrow Feshbach resonances provides predictions for the suppression of collisional decay as a function of the detuning from resonance, and we find excellent agreement with the experimental benchmarks provided by our $^6$Li-$^{40}$K system. 
We finally present model calculations for other Feshbach-resonant Fermi-Fermi systems, which are of interest for experiments in the near future. 
\end{abstract}

\maketitle

\section{Introduction}
The creation of weakly bound dimers near Feshbach resonances has led to major advances in the field of ultracold quantum gases \cite{Koehler2006poc,Ferlaino2008cmt,Chin2010fri}. 
Such Feshbach dimers have been the key to molecular Bose-Einstein condensation \cite{Jochim2003bec, Greiner2003eoa, Zwierlein2003oob} and to other applications, including the detection of atom pairs in strongly interacting fermionic superfluids \cite{Regal2004oor, Zwierlein2004cop} and in optical lattices \cite{Thalhammer2006llf, Winkler2006rba, Meinert2013qqi}. 
The weakly bound dimers can also serve as an excellent starting point for accessing the complex level structure of more deeply bound states \cite{Lang2008ctm} and, in particular, for creating ground-state molecules \cite{Lang2008utm, Ni2008ahp, Danzl2010auh, Takekoshi2014uds, Molony2014cou, Park2015udg, Guo2016coa}. 

For many applications, the stability of the dimers is of crucial importance. 
In particular, collisional quenching to lower vibrational states can release an amount of energy that greatly exceeds the depth of the trapping potential, and thus results in immediate losses from the stored sample. 
A special situation can arise for bosonic dimers formed in a two-component sample of fermionic atoms close to a Feshbach resonance. 
Here, a Pauli suppression effect \cite{Petrov2004wbm, Petrov2005spo, Petrov2005dmi} can dramatically reduce collisional losses to lower vibrational states, rendering such dimers exceptionally stable and facilitating their highly efficient evaporative cooling. 
This Pauli suppression effect has been observed and studied in strongly interacting spin mixtures of $^6$Li \cite{Cubizolles2003pol, Jochim2003pgo} and $^{40}$K \cite{Regal2004lom}, which both exhibit broad resonances. 
This has paved the way to spectacular achievements, such as molecular Bose-Einstein condensation \cite{Jochim2003bec, Greiner2003eoa, Zwierlein2003oob}, the experimental realization of the crossover to a Bardeen-Cooper-Schrieffer-type superfluid \cite{Giorgini2008tou}, and the exploration of the universal properties of resonantly interacting Fermi gases \cite{zwerger2011tbb}.

A central question for experiments exploring the many-body physics of fermionic mixtures is how far this suppression extends to mixtures of different species, featuring mass imbalance and narrow resonances.
Theoretical investigations have considered the important roles of the mass ratio \cite{Petrov2005dmi, Marcelis2008cpo} and of the resonance width \cite{Levinsen2011ada}.
The combination of $^6$Li and $^{40}$K atoms \cite{Taglieber2008qdt, Wille2008eau, Tiecke2010bfr, Naik2011fri, Trenkwalder2011heo} is the only Fermi-Fermi mixture with tunable interactions that has been experimentally realized so far and thus is the only available heteronuclear system that can provide experimental benchmarks. 
Dimers composed of $^6$Li and $^{40}$K atoms have been observed at LMU Munich in Ref.~\cite{Voigt2009uhf}, including preliminary lifetime studies, as well as in various recent experiments in Innsbruck \cite{Kohstall2012mac, Jag2014ooa}.  

In this Article, we present a joint experimental and theoretical investigation of the lifetime and decay properties of Feshbach dimers formed in a mixture of $^6$Li and $^{40}$K atoms. 
In Sec.~II, we describe the basic procedures for creating and investigating pure samples of Feshbach dimers and atom-dimer mixtures near a Feshbach resonance. 
In Sec.~III, we report on the measurements of spontaneous dissociation and of inelastic collisions in optically trapped dimer samples and in atom-dimer mixtures. 
Our results demonstrate the suppression of losses near the Feshbach resonance, but much weaker than that reported in Ref.~\cite{Voigt2009uhf} . 
In Sec.~IV, we present theoretical calculations based on the approach of Ref.~\cite{Levinsen2011ada} and find very good agreement with our observations. 
Finally, anticipating the creation of new mixtures, we present predictions for other Fermi-Fermi combinations with different mass ratios.

\section{Experimental Procedures}
\subsection{Feshbach resonances}
We employ two different Li-K interspecies Feshbach resonances (FRs).
The first resonance has been widely used in our previous work on Fermi-Fermi mixtures, including the observation of the hydrodynamic expansion of a strongly interacting mixture \cite{Trenkwalder2011heo}, the investigation of polarons \cite{Kohstall2012mac,Cetina2015doi,Cetina2016umb}, and the study of K-LiK atom-dimer interactions \cite{Jag2014ooa}. 
This resonance occurs near $155\,$G (width $0.88\,$G) with lithium in its lowest Zeeman sub-level Li$|1\rangle$ ($f = 1/2$, $m_f = +1/2$) and potassium in its third-lowest sub-level K$|3\rangle$ ($f=9/2$, $m_f = -5/2$).
The other resonance occurs near $158\,$G (width $0.14\,$G) with Li and K in their lowest-energy spin states Li$|1\rangle$ and K$|1\rangle$ ($f=9/2$, $m_f = -9/2$), respectively
We use the latter, narrower resonance for comparison as it has the advantage of an absence of any Li-K two-body losses.

The dependence of the Li-K $s$-wave scattering length $a$ on the magnetic field $B$ near a FR can be described by the standard expression $a(B) = a_\mathrm{bg}\left[ 1-\Delta/(B-B_0)\right]$ \cite{Chin2010fri} with the relevant background scattering length $a_\mathrm{bg}$, the width $\Delta$, and the resonance center $B_0$.
In Table~\ref{Table1} we summarize the values of these parameters for both resonances.
To fully characterize the FRs, we also present the differential magnetic moments $\delta\mu$ between the relevant open and closed channels.
From these parameters, we derive the length parameter $R^{*}=\hbar^2/(2 m_\mathrm{r} \Delta a_\mathrm{bg} \delta\mu)$ \cite{Petrov2004tbp}, characterizing the coupling strength between the open and the closed channel.
Here $m_\mathrm{r}$ represents the Li-K reduced mass.
The values for $a_\mathrm{bg}$ and $\Delta$ have been obtained from a coupled-channels calculation \cite{Naik2011fri}.
The values for $\delta\mu$ as well as $B_0$ for the Li$|1\rangle$-K$|3\rangle$ FR were experimentally determined, with very high accuracy, as described in Ref.~\cite{Cetina2015doi}.
For $\delta\mu$ near the Li$|1\rangle$-K$|1\rangle$ FR we use the data obtained from a coupled-channels calculation \cite{Naik2011fri} and for $B_0$ we use the value of an independent experimental determination \cite{UnPubB0}. 

\begin{table}
\centering
\begin{tabular}{|c|c|c|c|c|c|}
  \hline
  Channel 									& $B_0$ 			& $a_\mathrm{bg}$	& $\Delta$	& $\delta\mu/h$	& $R^*$			\\
														& (G)   			& ($a_0$)						& (G) 			& (MHz/G)			  & ($a_0$) 	\\
  \hline
  \hline
  Li$|1\rangle$K$|3\rangle$ & 154.708(2) 	& 63.0 							& 0.88 			& 2.35 					&  2\,650 	\\
    \hline
  Li$|1\rangle$K$|1\rangle$ & 157.530(3) 	& 65.0 							& 0.14 			& 2.3 					& 16\,500 	\\
  \hline
\end{tabular}
\caption{Parameters characterizing the two Feshbach resonances.
We summarize the values from Refs.~\cite{Cetina2015doi, Naik2011fri, UnPubB0} for the position $B_0$, background scattering length $a_\mathrm{bg}$, and width $\Delta$, as well as for the differential magnetic moment $\delta\mu$.
The values given for $B_0$ include a small shift ($9\,$mG) induced by the trapping-laser light \cite{Cetina2015doi}.}
\label{Table1}
\end{table}

\subsection{Sample preparation}
\label{subsec:SamplePreparation}
Our procedure to prepare Li$|1\rangle$K$|3\rangle$-dimer samples is essentially the same as the one described in Ref.~\cite{Jag2014ooa}.
To produce Li$|1\rangle$K$|1\rangle$ dimer samples, we slightly adapt this procedure to account for the narrower character of the FR. 
In both cases, the starting point for our experiments is an optically trapped and thermally equilibrated mixture of typically $10^5$ Li atoms and approximately $3\times 10^4$ K atoms at a temperature of $\sim 370\,$nK and at a magnetic field of $156.4\,$G. 
We reach these conditions by a preparation procedure described in detail in Ref.~\cite{Spiegelhalder2010aop}.
The cigar-shaped optical confinement of the atom mixture, realized by two 1064-nm laser-light beams intersecting at an angle of about $16^{\circ}$, is characterized by the radial and axial trap frequencies $\nu_\mathrm{r,\,K} = 420(10)\,$Hz and $\nu_\mathrm{a,\,K} = 55(2)\,$Hz for the K and $\nu_\mathrm{r,\,Li} = 600(10)\,$Hz and $\nu_\mathrm{a,\,Li} = 90(3)\,$Hz for the Li atoms.
At this stage, all Li atoms are in their lowest Zeeman sub-level Li$|1\rangle$ and all K atoms are in their second-lowest sub-level K$|2\rangle$ ($f=9/2$, $m_f = -7/2$).

The subsequent preparation steps differ depending on the Li-K spin-state combination from which the dimers are created.
To prepare for the creation of Li$|1\rangle$K$|3\rangle$ (Li$|1\rangle$K$|1\rangle$) dimers from these mixtures, we slowly ramp the magnetic field over $2\,$s to a value of $154.89\,$G ($157.565\,$G), approximately $180\,$mG ($35\,$mG) above the center of the FR.
Here, we transfer all K atoms into the K$|3\rangle$ (K$|1\rangle$) state by a radio-frequency rapid adiabatic passage. 

\begin{figure}
\centering
\includegraphics[clip,width=0.9\columnwidth]{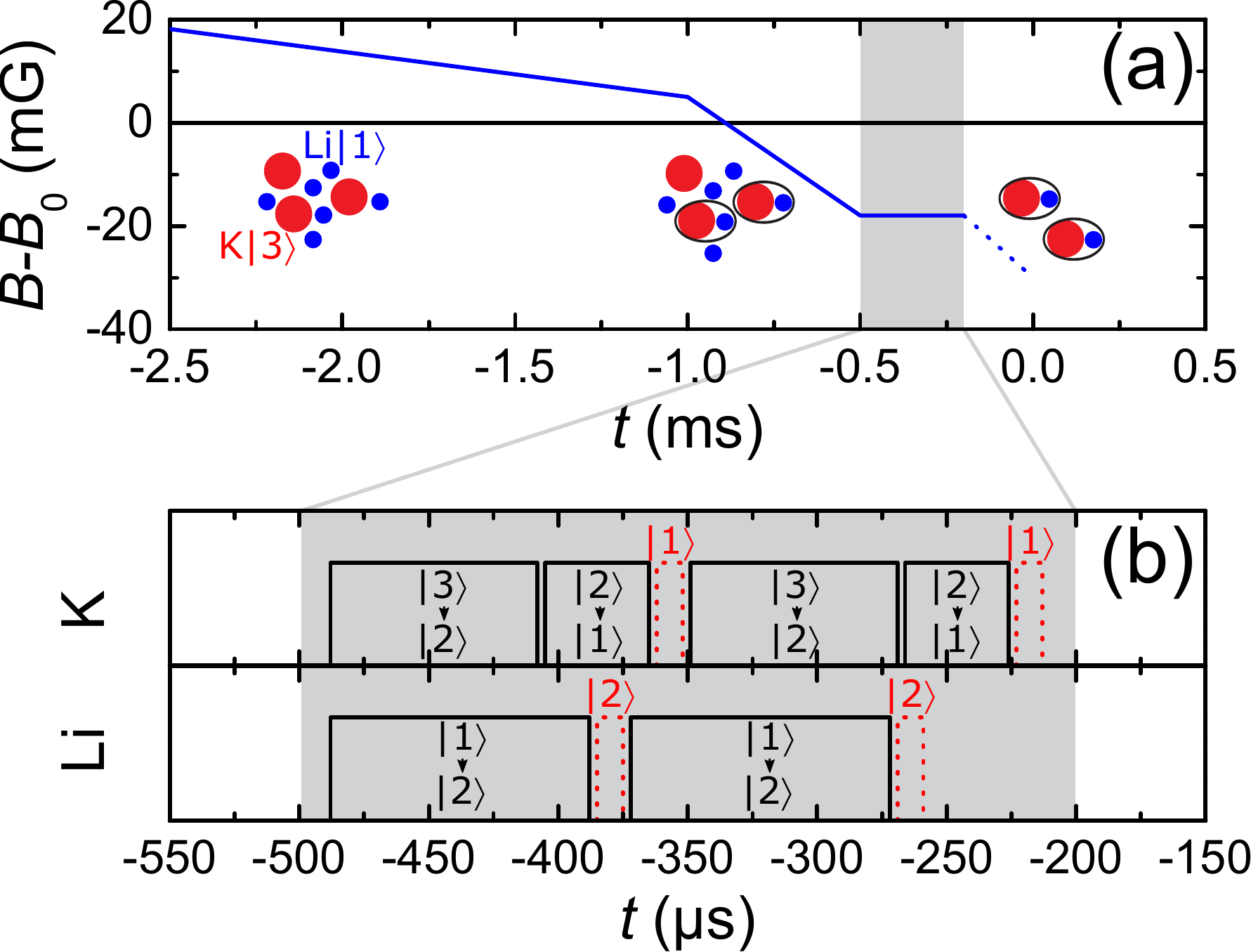}
\caption{(Color online) Schematic of the preparation of a pure Li$|1\rangle$K$|3\rangle$ dimer sample.
(a) Starting from the magnetic field $B = B_0+180\,$mG, we approach the resonance by a first 20-ms ramp to $B_0+5\,$mG (last $1.5\,$ms shown).
Then, we associate dimers by a quick ($0.5\,$ms) ramp across the FR to a magnetic field $B_0-20\,$mG. 
Here, within $0.3\,$ms (gray shaded), we remove unbound K and Li atoms from the sample.
After this cleaning procedure we reach the final magnetic field $B$, at which we perform the lifetime measurement, by a $200\,$-$\mu$s ramp (dotted line).
(b) The cleaning procedure for both Li and K consists of radio-frequency pulses (solid black), selectively transferring unbound atoms into another spin state, and successive removal of these atoms from the trap by a resonant laser-light pulse (dotted red).
This cleaning procedure is repeated one more time to increase the purity of the dimer sample.}
\label{Figure1}
\end{figure}

We then associate approximately $10^4$ LiK dimers by a Feshbach ramp \cite{Koehler2006poc, Chin2010fri}.
To associate dimers from the Li$|1\rangle$-K$|3\rangle$ mixture, we do this in two steps, as illustrated in Fig.~\ref{Figure1}(a).
In a first step, we ramp the magnetic field to $B_0+5\,$mG in $20\,$ms, which is sufficiently slow for the Li atoms to be attracted into the regions of high K density, increasing the density overlap between the two clouds.
This is followed by the second step, in which we quickly ramp the magnetic field to $B_0-20\,$mG in $0.5\,$ms.
For the Li$|1\rangle$-K$|1\rangle$ mixture, we associate the dimers by a single $2$-ms Feshbach ramp to a magnetic field $B = B_0 -16\,$mG, since here, at the much narrower FR, it is very hard to optimize a two-step ramping procedure.
Typical dimer numbers of Li$|1\rangle$K$|3\rangle$ samples are roughly $20\%$ larger than the typical numbers of Li$|1\rangle$K$|1\rangle$ samples.

To obtain pure dimer samples we apply cleaning sequences to remove unbound atoms.
For the Li$|1\rangle$K$|3\rangle$ samples, this sequence consists of a combination of radio-frequency (rf) and laser-light pulses; see Fig.~\ref{Figure1}(b). 
A 100-$\mu$s rf $\pi$-pulse selectively transfers the free Li atoms from the Li$|1\rangle$ state into the Li$|2\rangle$ state.
A subsequent $10$-$\mu$s laser pulse selectively removes the Li$|2\rangle$ atoms from the trap.
Simultaneous with this Li-cleaning procedure, we remove the unbound K atoms in a similar way. 
Applying two rf $\pi$-pulses with durations of $80\,\mu$s and $40\,\mu$s, we transfer the free K$|3\rangle$ atoms into the K$|1\rangle$ state, and successively remove them from the trap by applying a laser-light pulse resonant to the K$|1\rangle$ atoms.
As these cleaning procedures remove about $95\%$ of the free Li and K atoms, they are repeated one more time to clean the respective states more thoroughly.
For the Li$|1\rangle$K$|1\rangle$ samples, the Li cleaning is identical to the one explained above and the K cleaning is only slightly adapted. 
We revert the order of the 80-$\mu$s and 40-$\mu$s rf $\pi$-pulses to transfer the free K$|1\rangle$ atoms into the K$|3\rangle$ state and we then apply a laser pulse resonant to the K$|3\rangle$ atoms to remove them from the trap. 
After the cleaning procedure, we quickly, within $200\,\mu$s, ramp the magnetic field to its variable final value, at which we then perform the measurements.

\subsection{Dimer detection and dimer-temperature determination}
We determine the LiK-dimer numbers from absorption images of Li and K atoms after dissociation of the dimers into Li-K pairs by a reverse Feshbach ramp \cite{Koehler2006poc, Chin2010fri}.
For both resonances we ramp the magnetic field $B$ up to a value $\geq B_0+50\,$mG within $10\,\mu$s. 
After an additional wait time of a few $10\,\mu$s, we simultaneously take absorption images of the Li and the K cloud, from which we determine the numbers of Li and K atoms.
In some measurements, we detected only the number of Li atoms remaining after the reverse Feshbach ramp.

The temperature of the dimers is determined from Gaussian fits to absorption images of the clouds after a time-of-flight expansion duration of $t_\mathrm{TOF} = 4\,$ms.
The procedure is discussed in detail in Ref.~\cite{Jag2014ooa}.
From the measured radial width $\sigma_\mathrm{r}$, we obtain the dimer temperature $T_\mathrm{D}$ from $k_\mathrm{B}T_\mathrm{D} = m_\mathrm{D} (\sigma_\mathrm{r}/t_\mathrm{TOF})^{2}$, where $m_\mathrm{D} = m_\mathrm{Li} +m_\mathrm{K}$ is the mass of a Li-K dimer.
Typically, the temperatures of our dimer samples are about $T_\mathrm{D} = 550\,$nK. 
This corresponds to peak phase-space densities of about $0.1$ for typical dimer number densities in our samples. 
We explain the increased temperature of our dimer cloud compared to the temperature prior to the dimer association ($370\,$nK) by heating and collective excitations caused by our preparation procedure \cite{Jag2014ooa}.

\section{Measurements of Dimer Decay}
In this Section, we present measurements characterizing various processes that lead to losses of LiK dimers. 
In Sec.~\ref{subsec:Spont}, we first discuss spontaneous dissociation, which, being a one-body process, can also occur in very dilute samples. 
In Secs.~\ref{subsec:DiDiColl} and \ref{subsec:AtDiColl}, we then present our experimental results on dimer-dimer collisions and atom-dimer collisions, which, as two-body processes, limit the lifetime of dense samples.

\subsection{Spontaneous dissociation}
\label{subsec:Spont}
A dimer created from an atom pair with at least one atom in an excited Zeeman state can spontaneously decay via processes mediated by the coupling between the two spins \cite{Chin2010fri}.
Such decay has previously been studied theoretically and experimentally for the case of $^{85}$Rb$_2$ molecules \cite{Koehler2005sdo, Thompson2005sdo}.
Our Li$|1\rangle$K$|3\rangle$ dimers are also subject to this decay process, in contrast to the Li$|1\rangle$K$|1\rangle$ combination.
The spontaneous decay of Li$|1\rangle$K$|3\rangle$ dimers has been theoretically investigated in detail in Ref.~\cite{Naik2011fri}, where predictions for the lifetimes of the dimers were obtained from coupled-channels calculations.

\begin{figure}
\centering
\includegraphics[clip,width=0.98\columnwidth]{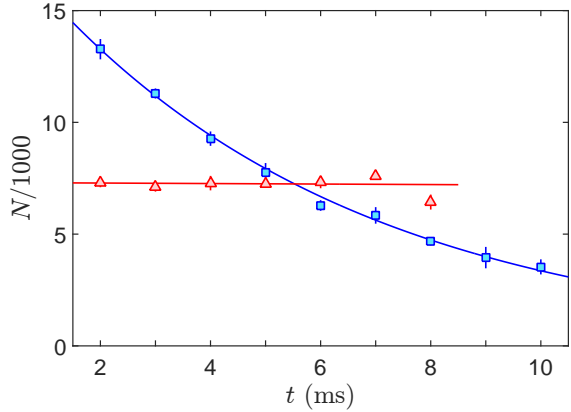}
\caption{(Color online) Comparison of the dimer number evolution near the Li$|1\rangle$-K$|3\rangle$ and the Li$|1\rangle$-K$|1\rangle$ FR.
The blue squares show a typical decay curve of a Li$|1\rangle$K$|3\rangle$-dimer sample at $B = B_0-296\,$mG.
Fitting an exponential decay to the data yields the 1/e-lifetime $\tau = 5.8(4)\,$ms.
The fit is represented by the blue solid line.
The results from similar measurements with a Li$|1\rangle$K$|1\rangle$-dimer sample at a magnetic detuning of $-75\,$mG from the respective resonance center, are shown as the red triangles.
Here, we observe the dimer number to remain essentially constant.
A fit of an exponential decay to the data (red solid line) is consistent with infinite lifetime.
The error bars represent $1\sigma$ uncertainties; in some cases they are smaller than the symbol size.}
\label{Figure2}
\end{figure}

We experimentally investigate the lifetime of Li$|1\rangle$K$|3\rangle$ with respect to spontaneous decay using dimer samples with a very low number density, so that density-dependent collisional losses do not play a significant role.
We realize such dilute dimer samples by allowing the optically trapped dimer cloud to expand after switching off the trap.
After a variable expansion time $t$, we determine the molecule number in the sample.
Note that the 1064-nm light induces a shift of the FR center $B_0$, as described in Ref.~\cite{Cetina2015doi}.
When the optical trap is off, the FR center $B_0$ of the Li$|1\rangle$-K$|3\rangle$ resonance is found at $154.699\,$G, i.e. $9\,$mG lower than in the trap (Table~\ref{Table1}). 
For the Li$|1\rangle$-K$|1\rangle$ channel we assume the same small shift.

In Fig.~\ref{Figure2} we show a typical decay curve of a Li$|1\rangle$K$|3\rangle$-dimer sample, recorded at a magnetic detuning $B -B_0 = - 296\,$mG (blue squares).
For our analysis, we only consider data obtained for $t\geq 1.5\,$ms, where the mean dimer number density in the sample is below $5\times 10^{10}/\mathrm{cm}^3$, low enough for collisional losses to play a negligible role.
To these data we fit a simple exponential decay, $N_0 \exp{(-t/\tau)}$, with the initial dimer number $N_0$ and the lifetime $\tau$ as free parameters.
For the specific example of Fig.~\ref{Figure2}, this procedure yields $\tau = 5.8(4)\,$ms and the fit result is shown as the blue solid line. 

For comparison, we also show the evolution of the number of Li$|1\rangle$K$|1\rangle$ dimers recorded $75\,$mG below the center of the Li$|1\rangle$-K$|1\rangle$ resonance (red triangles).
Here, the spontaneous decay mechanism is absent.
Indeed, we observe an essentially constant number of Li$|1\rangle$K$|1\rangle$ dimers, with the fit yielding the decay rate $1/\tau = 0.008(7)\,$s$^{-1}$. 
This result is essentially consistent with an infinite lifetime and, at a $95\%$ confidence level, provides a lower bound of $50\,$ms.

\begin{figure}
\centering
\includegraphics[clip,width=0.98\columnwidth]{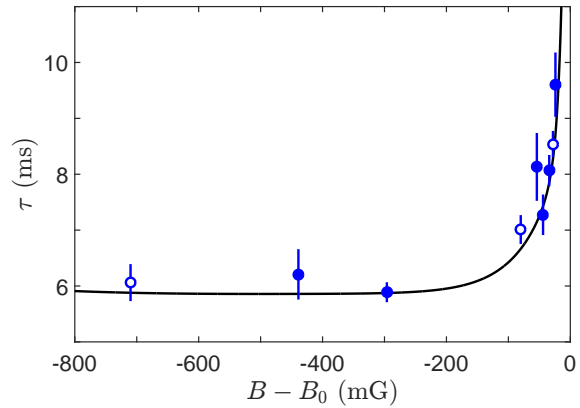}
\caption{(Color online) Lifetime of dimers against spontaneous decay near the Li$|1\rangle$-K$|3\rangle$ FR.
The data points show the experimental results and the black solid line represents the theoretical prediction from Ref.~\cite{Naik2011fri}.
While the filled symbols are obtained from decay curves, where both the Li and the K component have been imaged after dissociation, the open symbols are based on detecting K alone.
The error bars represent the 1$\sigma$ fit uncertainties.}
\label{Figure3}
\end{figure}

In Fig.~\ref{Figure3}, the blue circles show the measured lifetimes of the dimers with respect to spontaneous decay over a wide range of magnetic detunings $B-B_0$.
Comparing our experimental results to the predictions from Ref.~\cite{Naik2011fri} (black solid line), we find an excellent agreement over the whole magnetic field range investigated.
While for magnetic detunings of around a few hundred mG the lifetime is about $6\,$ms, we in particular confirm the predicted substantial increase near the FR, where we determine lifetimes approaching $10\,$ms.
This increase can be attributed to the increasing halo character of the dimer wave function as the FR is approached.
This leads to a decreased probability to find a pair of Li and K atoms within the short range where the spin coupling occurs \cite{Naik2011fri}.
Our measurements of the lifetime of the Li$|1\rangle$K$|3\rangle$ dimers in {\em dilute} samples can be fully understood in terms of spontaneous dissociation.

\subsection{Dimer-dimer collisions}
\label{subsec:DiDiColl}
In a second series of experiments, we investigate the collisional decay of a trapped dimer cloud.
In collisions with other dimers our shallowly bound dimers can relax into more deeply bound states.
The binding energy that is released in this process is much larger than the depth of the trapping potential, and thus the relaxation products are always lost from the trap.
This two-body decay occurs at a rate $\beta_\mathrm{D} n$, which is equal to the product of the dimer-dimer two-body loss-rate coefficient $\beta_\mathrm{D}$ and the dimer number density $n$.

\begin{figure}
\centering
\includegraphics[clip,width=0.98\columnwidth]{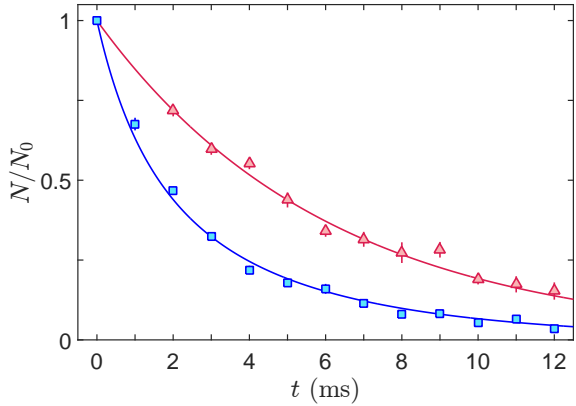}
\caption{(Color online) Comparison of the decay of a trapped and an expanding Li$|1\rangle$K$|3\rangle$-dimer sample. 
The blue squares show the measured dimer number in a trapped sample versus hold time $t$ in the trap.
The red triangles show the dimer number determined in a dilute, expanding sample, $1.5\,$ms after release from the trap.
The blue and red lines correspond to the fit of our model to the data without and with two-body decay (see text).
To enable a direct comparison, the experimental data are normalized to the initial dimer number $N_0=13000$ (15300) obtained from the fit to the data acquired from the trapped (expanding) sample.
The error bars represent $1\sigma$ uncertainties; in some cases they are smaller than the symbol size.}
\label{Figure4}
\end{figure}

To experimentally determine the rate coefficient $\beta_\mathrm{D}$ for these collisional decay processes, we investigate the decay of a trapped sample of dimers.
The initial number of typically $N_0 = 1.3\times10^{4}$ dimers corresponds to an initial number density $N_0/V_\mathrm{eff}$ of about $1\times 10^{12}/$cm$^{3}$, where $V_\mathrm{eff} = [(4\pi k_\mathrm{B} T_\mathrm{D})/(m_\mathrm{D}\bar{\omega}_\mathrm{D}^2)]^{3/2}$ is the effective volume of a thermalized sample, and $\bar{\omega}_\mathrm{D}= 2 \pi (\nu_\mathrm{r,\,D}^2 \nu_\mathrm{a,\,D})^{1/3} \approx 2 \pi \times 230\,$Hz is the mean dimer trapping frequency \cite{Note0}. 
After a hold time $t$ at a magnetic field $B$ we measure the number of dimers, $N(t)$, remaining in the sample.
In Fig.~\ref{Figure4} we show an example for a decay curve obtained at a magnetic detuning of $-710\,$mG from the Li$|1\rangle$-K$|3\rangle$ FR (blue squares).

We model the decay with the common loss-rate equation
\begin{equation}
\dot{N}/N = -1/\tau-(\beta_\mathrm{D}/V_\mathrm{eff}) N.
\label{Eq:DimerLossDEQ}
\end{equation}
Under the assumption that the sample remains in thermal equilibrium at the initial temperature $T_\mathrm{D}$, this differential equation has the solution
\begin{equation}
N(t) = \frac{N_0 \exp(-t/\tau)}{1 + \frac{\beta_\mathrm{D}}{V_\mathrm{eff}}N_0 \tau \left[1-\exp(-t/\tau)\right]}.
\label{Eq:DimerLossSolution}
\end{equation}
We fit Eq.~(\ref{Eq:DimerLossSolution}) to the experimental data to extract the loss-rate coefficient $\beta_\mathrm{D}$.
While $\beta_\mathrm{D}$ and $N_0$ are free parameters, we fix $\tau$ to the corresponding theoretical value, which was verified in the independent measurements presented before.
For the data of Fig.~\ref{Figure4}, the fit result is shown as the blue solid line.
For comparison, we also show the decay curve of a dilute dimer sample, where collisional loss is absent (red triangles), together with the result of a fit of a simple exponential decay to this data (red line).
Our measurements show that, under typical experimental conditions, the collisional relaxation and the spontaneous dissociation give similar contributions to the total decay of the trapped dimer sample. 

The given values for the loss coefficients $\beta_\mathrm{D}$ are subject to a systematic error arising from an uncertainty in the dimer number density.
We estimate a combined systematic error of about 40\%, arising from largely uncorrelated uncertainties of 25\%, 7\%, and 20\% in the dimer number, the dimer trapping frequencies, and the dimer temperature, respectively.
Furthermore, by assuming a constant temperature $T_\mathrm{D}$ of the decaying dimer sample, and thus a constant $V_\mathrm{eff}$ in Eq.~(\ref{Eq:DimerLossDEQ}), we neglect a small effect of anti-evaporation heating \cite{Weber2003tbr}.
We have checked that including the latter into our analysis would lead to slightly larger values for $\beta_\mathrm{D}$. 
We found this correction to stay well below $15\%$.

\begin{figure}
\centering
\includegraphics[clip,width=0.98\columnwidth]{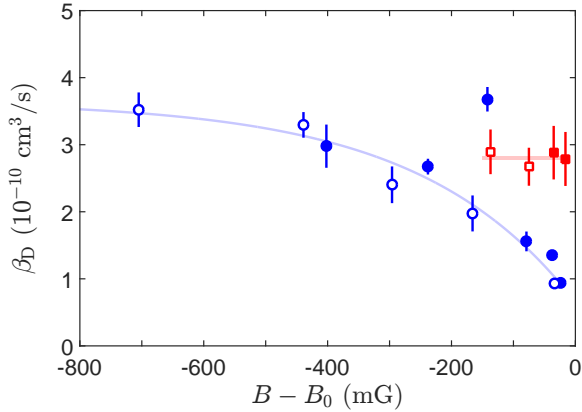}
\caption{(Color online) Measured loss-rate coefficient for inelastic dimer-dimer collisions as a function of magnetic detuning.
The blue circles (red squares) correspond to the experimental results obtained with samples of Li$|1\rangle$K$|3\rangle$ (Li$|1\rangle$K$|1\rangle$) dimers.
The filled symbols correspond to results we obtained when determining the molecule number from both Li and K absorption images.
Open circles (squares) represent fit results based on analyzing Li (K) images alone.
The error bars represent the 1$\sigma$ fit uncertainties; in some cases they are smaller than the symbol size.
We show the light blue and a light red line as guides to the eye.}
\label{Figure5}
\end{figure}

We determine the values for the loss coefficient $\beta_\mathrm{D}$ at various magnetic detunings. 
Our experimental results, obtained with Li$|1\rangle$K$|3\rangle$ (Li$|1\rangle$K$|1\rangle$) dimer samples, are shown in Fig.~\ref{Figure5} as the blue circles (red squares). 
For the Li$|1\rangle$K$|1\rangle$ dimer samples we obtain values for the loss-rate coefficient $\beta_\mathrm{D}$ of roughly $3 \times 10^{-10}$cm$^3$/s without significant dependence on the magnetic detuning. 
Also for the Li$|1\rangle$K$|3\rangle$ dimer samples we obtain roughly the same value for detunings $B-B_0 \lesssim -400\,$mG.
At these large magnetic detunings, the Feshbach molecules have a very small admixture of the entrance channel and are thus strongly closed-channel dominated.
As we discuss in more detail in Sec.~IV, the decay of such molecules is largely independent of the exact state they are in \cite{Staanum2006eio,Zahzam2006amc,Gao2010umf,Julienne2011uuc,Quenemer2011uiu}, which explains why the measurements for both FRs at large detunings result in nearly the same values.

As the Li$|1\rangle$-K$|3\rangle$ resonance is approached, our experimental results (with the exception of one clear outlier 
\cite{Note1}) show a reduction of collisional losses, which we interpret in terms of the Pauli-suppression effect. 
For our data points closest to resonance (about $-30\,$mG detuning), this suppression effect amounts to more than a factor of three. 
Note that measurements closer to resonance are prevented by the onset of collisional dissociation \cite{Jochim2003pgo}.

\subsection{Atom-dimer collisions}
\label{subsec:AtDiColl}
In another set of experiments, we study the decay of dimers arising from their collisions with Li atoms in a mixture of LiK dimers and Li atoms.
Such decay occurs at a rate $\beta_\mathrm{LiD} n_\mathrm{Li}$, equal to the product of the Li atom dimer loss coefficient $\beta_\mathrm{LiD}$ and the Li density $n_\mathrm{Li}$.
The measurement of atom-dimer collisions is challenging because the corresponding decay has to be distinguished from both the spontaneous decay and the dimer-dimer collisional decay.

We realize mixtures of Li atoms and LiK dimers by adapting the preparation procedure presented in Sec.~\ref{subsec:SamplePreparation}.
Here we start with the lithium component in a nearly balanced spin mixture of Li$|1\rangle$ and Li$|2\rangle$. 
The Feshbach ramp then produces a mixture of Li-K dimers, some remaining Li$|1\rangle$ atoms, and the unaffected Li$|2\rangle$ atoms. 
Then, at $B = B_0-20\,$mG, we apply only one radio-frequency $\pi$ pulse, which exchanges the populations of the Li$|1\rangle$ and Li$|2\rangle$ states.
We subsequently remove the Li$|2\rangle$ atoms from the trap using a laser-light pulse.
All other preparation steps, in particular the K spin state cleaning, remain as described in Sec.~\ref{subsec:SamplePreparation}.
After this procedure, the number density distribution of the Li atoms in the trap, $n_\mathrm{Li}$, can be well approximated by the density of a noninteracting Fermi gas at a temperature equal to the initial Li temperature.
Typically, we obtain samples of $\sim 9\times10^3$ dimers and a mildly degenerate Fermi sea of $\sim 6\times10^4$ Li atoms at a temperature that is about 55\% of the Fermi temperature.
This corresponds to a mean dimer density of $6\times10^{11}/\mathrm{cm}^3$ and a Li density averaged over the dimer distribution \cite{Cetina2016umb}, $\langle n_\mathrm{Li} \rangle$, of about $1.5\times10^{12}/\mathrm{cm}^3$.

To experimentally determine the rate coefficient $\beta_\mathrm{LiD}$, we again investigate the decay of dimers from our sample.
We ramp the magnetic field to a desired value $B$ and, after a variable hold time $t$, we measure the number of dimers, $N$, remaining in the sample.
For each decay curve in the atom-dimer mixture we record a corresponding reference curve in a pure dimer sample. 
These reference measurements, which independently provide the dimer-dimer loss coefficient $\beta_\mathrm{D}$, are the ones that we have presented in the preceding section. 
To minimize systematic errors resulting from long-term drifts of the experiment, the measurements in the atom-dimer mixtures and the pure dimer samples are carried out in alternating order.

We model the decay of dimers with a simple extension of the decay model from the previous section.
Our Li sample is much larger than the dimer sample, such that losses from the Li sample can be neglected.
In this case, the Li sample represents a constant-density bath and the loss of dimers arising from Li atom dimer collisions appears as a one-body loss, which we include into our model by adding $-\beta_\mathrm{LiD} \langle n_\mathrm{Li} \rangle$ to the right-hand side of Eq.(\ref{Eq:DimerLossDEQ}).
Under these assumptions, the solution of our model is given by substituting $\tau^{-1}$ with $\beta_\mathrm{LiD} \langle n_\mathrm{Li} \rangle + \tau^{-1}$ in Eq.(\ref{Eq:DimerLossSolution}).
We fit this solution to our experimental data to determine the Li atom dimer loss coefficient $\beta_\mathrm{LiD}$.
For the fit, we fix $\tau$ to the corresponding theoretical value and the decay coefficient $\beta_\mathrm{D}$ to the value we determined in the corresponding reference measurement on a pure dimer sample.

\begin{figure}
\centering
\includegraphics[clip,width=0.98\columnwidth]{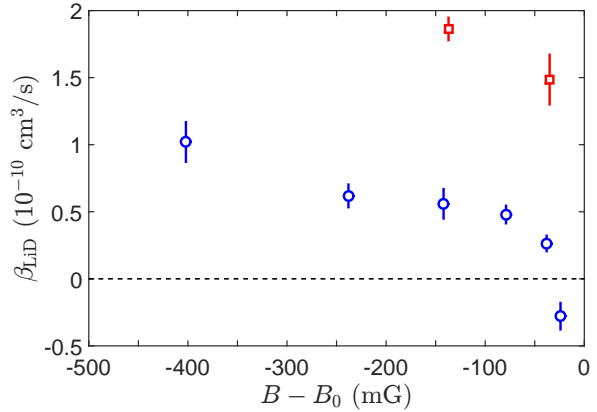}
\caption{(Color online) Measured loss-rate coefficient for inelastic Li atom dimer collisions as a function of the magnetic detuning.
The blue circles (red squares) correspond to the experimental results obtained with samples of Li$|1\rangle$K$|3\rangle$ (Li$|1\rangle$K$|1\rangle$) dimers co-trapped with Li$|1\rangle$ atoms.
In these experiments, the dimer number was determined from the K absorption images only. 
The error bars include the combined fit uncertainties (see text).}
\label{Figure6}
\end{figure}

In Fig.~\ref{Figure6}, we show our results for the Li atom dimer loss coefficient $\beta_\mathrm{LiD}$ at various magnetic detunings.
The blue circles (red squares) correspond to data acquired with a Li$|1\rangle$K$|3\rangle$ (Li$|1\rangle$K$|1\rangle$) dimer sample.
The error bars reflect the 1$\sigma$ fit uncertainty of $\beta_\mathrm{LiD}$ as well as the contribution arising from the uncertainty in our determination of $\beta_\mathrm{D}$.
We obtain atom-dimer loss-rate coefficients of roughly $1.5\times10^{-10}$cm$^3/$s near the Li$|1\rangle$K$|1\rangle$ FR, where the molecules have closed-channel character. 
The data obtained with Li$|1\rangle$K$|3\rangle$ dimers show a suppression of atom-dimer collisional losses, which becomes stronger as we approach the FR and the open-channel fraction of the dimers increases.
The data point at a magnetic detuning of about $-40\,$mG already shows a suppression by a factor of roughly five.
From the data point at $-24\,$mG, we determine a negative loss coefficient.
We speculate that this unphysical result is due to the repulsive mean-field interaction between the dimers and the Li atoms, effectively increasing the cloud sizes and therefore decreasing the mean densities of the dimers and the Li atoms.
Such an effect is beyond the assumptions of the model underlying our data analysis and can therefore produce unphysical results. 
We estimate that all other values, taken at larger detunings, do not suffer from such interaction effects.
The observed suppression of atom-dimer collisional losses appears very similar to the effect observed in dimer-dimer decay, and can also be attributed to the Pauli suppression effect. 

\subsection{Summary of experimental results and comparison with previous work}
\label{subsec:SummaryExp}
Our experimental results characterize three different loss processes of $^6$Li$^{40}$K dimers close to a Feshbach resonance. 
Spontaneous dissociation was identified as a density-independent one-body loss mechanism.
This process is possible for Feshbach molecules composed of atoms that are not in the energetically lowest combination of spin states. 
For the case of the 155-G resonance in the Li$|1\rangle$-K$|3\rangle$ mixture, this limits lifetimes to values below $10\,$ms for typical experimental conditions. 
We have also investigated losses due to inelastic collisions in pure dimer samples, and obtained loss-rate coefficients of typically $3 \times 10^{-10}\,$cm$^3$/s. 
At the typical densities of near-degenerate molecular samples, the corresponding loss rate is similar to the effect of spontaneous dissociation. 
Additional losses occur in atom-dimer mixtures, as we have shown for the example of free excess Li atoms.

Very close to the resonance center, in a roughly 100-mG wide range, we observe a suppression of loss in both spontaneous and collisional decays. 
In the former case, the suppression is a direct consequence of the halo character of the molecular wavefunction \cite{Koehler2005sdo,Thompson2005sdo,Naik2011fri}. 
In the latter case, the suppression effect can be attributed to Pauli blocking \cite{Petrov2004wbm, Petrov2005dmi}, as we will discuss in more detail in Sec. IV. 
For the specific FR employed in the Li-K mixture, the suppression of loss only leads to an increase of dimer lifetimes by up to a factor of three.

Weakly bound $^6$Li$^{40}$K dimers have been created in previous work by the Munich-Singapore group \cite{Voigt2009uhf, Costa2010swi}, who investigated lifetime properties without distinguishing between different processes. 
Below the FR center, their observations are consistent with our results and can be understood as a combination of spontaneous and collisional losses. 
Above the FR center, in a 100-mG wide range, the Munich-Singapore work reports on molecules with lifetimes of more than $100\,$ms \cite{Voigt2009uhf}.
These long lifetimes were later interpreted in terms of a many-body effect \cite{Costa2010swi}. 
In our present work, using the same FR, we do not observe any molecules above resonance. 
In our previous work \cite{Kohstall2012mac}, with K impurities in a degenerate Li Fermi sea, we indeed observed indications of many-body pairs above resonance, though restricted to a narrow, less than 20-mG wide magnetic-field range. 
For the 155-G FR in the Li-K mixture, we cannot confirm the existence of long-lived ($\approx 100\,$ms) molecules.

\section{Theoretical Analysis of Relaxation Rates}
\label{sec:theory}
In this section we present a theoretical description of atom-dimer and dimer-dimer relaxation processes near a narrow resonance. 
The model has been introduced in Ref.~\cite{Levinsen2011ada} for characterizing atom-dimer and a subset of the dimer-dimer inelastic channels. 
In Sec.~\ref{sec:theory}A, we extend the discussion to all relevant dimer-dimer relaxation processes.
In Sec.~\ref{sec:theory}B, we then compare the theory with our experimental results and find a very good agreement.

\subsection{Theoretical model}
The collisional decay requires at least three atoms to approach each other to within distances comparable to the van der Waals range $R_e$ of the interatomic interactions (we call this the ``recombination region''). 
For the Li-K interaction, the van der Waals range takes the value $R_e=40.8 a_0$ \cite{Naik2011fri}. 
In relaxation channels involving three atoms, two atoms form a deeply bound state and the large binding energy is released as kinetic energy. 
As the central point of our model, the probability of such a relaxation event may be calculated within a theory that only describes the few-body kinematics at length scales greatly exceeding $R_e$, the short-range relaxation physics being characterized by the loss-rate constant for collisions of atoms with closed-channel (cc) molecules. 
One can show~\cite{Levinsen2011ada} that in the narrow-resonance limit, $R_e \ll R^*,a$, three atoms enter the recombination region predominantly as a free atom and a cc molecule rather than three free (open-channel) atoms. 
Thus, the recombination process is microscopically the relaxation in collisions of cc molecules with atoms. 
We assume that the corresponding interaction is not resonant and is characterized by a coupling constant $-i\Delta_\text{AD} \sim-i\hbar^2 R_e/m_\text{AD}$, where $m_\text{AD}$ is the atom-molecule reduced mass. 
The atom-cc molecule relaxation rate constant equals $\beta_{\rm AD}^{(0)}=2\Delta_\text{AD}/\hbar$. 
This relation can be derived by relating the lifetime of the atom and cc molecule to the imaginary part of their mean-field interaction energy shift in unit volume. 
The ``bare'' relaxation rate constant $\beta_{\rm AD}^{(0)}$ is an external parameter of our theory. 

\begin{figure}
\centering
\includegraphics[clip,width=0.98\columnwidth]{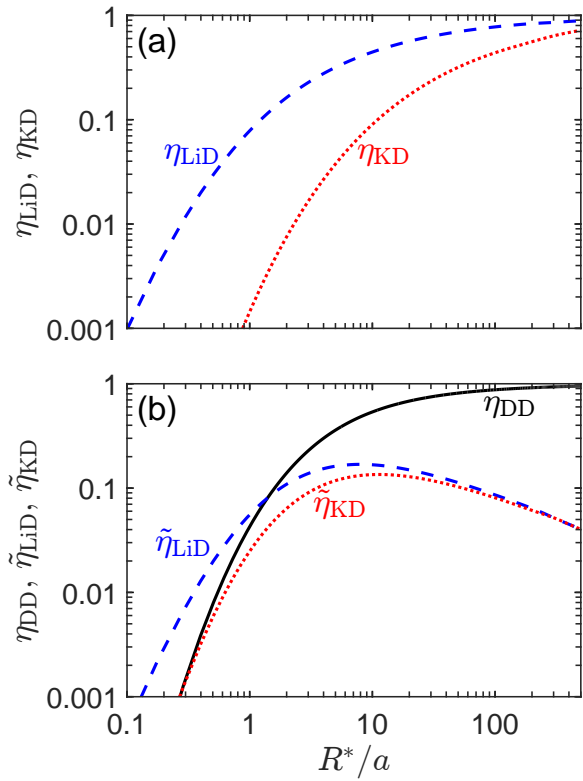}
\caption{(Color online) Suppression functions for relaxation in (a)~atom-dimer and (b) dimer-dimer collisions.}
\label{fig:relaxtheory1}
\end{figure}

In our approach, the atom-dimer relaxation rate constant $\beta_{\rm AD}$ factorizes into the product
\begin{equation}\label{beta_AD}
\beta_{{\rm AD}}=\beta_{\rm AD}^{(0)}\eta_{\rm AD}(R^*/a),
\end{equation}
where the dependence on the short-range physics is fully absorbed into $\beta_{\rm AD}^{(0)}$ and the long-range kinematics enters as the probability of finding an atom and cc molecule in the recombination region. 
This probability can be interpreted as the reduction of atom-dimer relaxation at finite $R^*/a$ and we refer to it as the ``suppression function'', $\eta_{\rm AD}(R^*/a)$. 
It depends only on $R^*/a$ and is proportional to the squared modulus of the atom-dimer wave function calculated under the assumption $\Delta_{\rm AD}=0$. 
The task of computing the normalization integral for this wave function, which is quite complex (particularly, in the four-body case discussed below) and contains closed- and open-channel components, can be avoided by using an equivalent diagrammatic formulation of the problem, see Ref.~\cite{Levinsen2011ada} where this approach was used for K-(K-Li) collisions. 
Namely, we calculate the atom-dimer scattering length $a_\text{AD}$ perturbatively to first order in $\Delta_{\rm AD}$ and deduce the atom-dimer relaxation rate constant from ${\rm Im}(a_{\rm AD})$.
The suppression function $\eta_{\rm AD}(R^*/a)$ is shown in Fig.~\ref{fig:relaxtheory1}(a) for the case of a light atom (A=Li) and for a heavy atom (A=K).
It is seen how the relaxation can be substantially reduced for $R^*/a \lesssim 1$, and that the suppression is stronger in the collision of the heavy K atom with the dimer.

In molecule-molecule collisions there are three possible relaxation channels: 
the Li-cc molecule, K-cc molecule, and cc molecule-cc molecule relaxation (we call it four-atom mechanism). 
The latter originates from inelastic collisions of cc molecules with each other involving no free atoms. 
This configuration dominates the four-body wave function when all four atoms are at distances smaller than $R^*$. 
We assume that three coupling constants $\Delta_{\rm LiD}$, $\Delta_{\rm KD}$, and $\Delta_{\rm DD}$ are proportional to the corresponding van der Waals ranges which are small compared to $R^*$ and $a$. 
This allows us to treat these interactions independently as first-order perturbations on top of the zero-order solution -- the properly normalized four-body wave function calculated for $\Delta_{\rm LiD}=\Delta_{\rm KD}=\Delta_{\rm DD}=0$. 
The contribution of a relaxation channel, say Li-cc molecule channel, to the total dimer-dimer relaxation rate constant $\beta_{\rm D}$ is the product of $\beta_{\rm LiD}^{(0)}=2\Delta_{\rm LiD}/\hbar$ and the probability $\tilde{\eta}_{\rm LiD}(R^*/a)$ to find a Li atom close to a cc molecule in dimer-dimer collisions. 
The quantity $\tilde{\eta}_{\rm LiD}(R^*/a)$, which is not to be confused with $\eta_{\rm LiD}$ defined for atom-dimer collisions, can in principle be calculated from the squared modulus of the zero-order four-body wave function by integrating it over the coordinates of one K atom (for the Li-cc molecule loss channel) and taking into account combinatorial factors (choice between two Li atoms). 
However, as in the atom-dimer case, we calculate the dimer-dimer scattering length $a_\text{DD}$ to first order in $\Delta_{\rm LiD}$, $\Delta_{\rm KD}$, and $\Delta_{\rm DD}$ and deduce the dimer-dimer relaxation rate constant from ${\rm Im}(a_{\rm DD})$. 
The total relaxation rate constant in dimer-dimer collisions is written in the form
\begin{equation}\label{beta_DD}
\beta_{{\rm D}}=\beta_{\rm LiD}^{(0)}\tilde{\eta}_{\rm LiD}(R^*/a)+\beta_{\rm KD}^{(0)}\tilde{\eta}_{\rm KD}(R^*/a)+\beta_{\rm D}^{(0)}\eta_{\rm DD}(R^*/a).
\end{equation}
The function $\eta_{\rm DD}(R^*/a)$ has been computed in Ref.~\cite{Levinsen2011ada}. 
Here we calculate $\tilde{\eta}_{\rm LiD}(R^*/a)$ and $\tilde{\eta}_{\rm KD}(R^*/a)$, as described in the Appendix. 
We show these functions together in Fig.~\ref{fig:relaxtheory1}(b); again we see how collisional losses can be strongly suppressed for $R^*/a \lesssim 1$, and that relaxation losses originating from light atoms and cc molecules are more important than those from heavy atoms and cc molecules.

Let us now discuss the limiting case of large detuning, $R^*\gg a$. 
Neglecting the open-channel population we obtain $\eta_{\rm DD}(R^*/a\rightarrow \infty)=1$ and $\tilde{\eta}_{\rm AD}(R^*/a\rightarrow \infty)=0$, where A=Li,K. 
Note that $\tilde{\eta}_{\rm AD}$ is not equal to $\eta_{\rm AD}$, which is defined for atom-molecule collisions and tends to 1 in the large $R^*/a$ limit. 
As expected, these results mean that $\beta_{{\rm D}}(R^*/a\rightarrow \infty)=\beta_{\rm D}^{(0)}$ and $\beta_{{\rm AD}}(R^*/a\rightarrow \infty)=\beta_{\rm AD}^{(0)}$. 
For large but finite $R^*/a$ we can perturbatively take into account the probability to be in the open channel $P_{\rm open}\approx \sqrt{a/4R^*}\ll 1$ arriving at $\tilde{\eta}_{\rm AD}\approx 2 P_{\rm open}\approx \sqrt{a/R^*}$ and $\eta_{\rm DD}\approx P_{\rm closed}^2\approx 1-\sqrt{a/R^*}$. 

In the opposite limit of small detuning, $R^*/a\ll 1$, the feature of particular interest is the suppression of collisional relaxation which arises from the large open-channel probability combined with Pauli suppression: 
The inelastic process requires at least three atoms -- of which two are identical fermions -- to approach each other. 
More precisely, the Pauli suppression mechanism is efficient at distances (hyperradii) $R^*\ll r \ll a$, which is called the universal region, where the atoms behave as free (open-channel) atoms. 
At shorter distances the three-atom configuration changes to the atom plus cc molecule one which is insensitive to the statistical suppression. 
This argument applies to the atom-cc molecule relaxation mechanism in both atom-dimer and dimer-dimer collisions, thus suppressed by the factor
\begin{align}
\eta_{\rm AD}\propto \tilde{\eta}_{\rm AD}\propto (R^*/a)^{2\nu_s+1}.
\label{eq:etaAD}
\end{align}
The exponent $\nu_s$ characterizes the three-body wave function in the universal region and depends on the masses, quantum statistics of atoms, and the total angular momentum (the subscript $s$ means $l=0$) \cite{Note2}. 
For the relevant cases of Li (K) atoms scattering on LiK dimers we have $\nu_s\approx1.01$ ($\nu_s\approx2.02$), respectively~\cite{Levinsen2011ada}. The onset of the power-law suppression can be seen in Fig.~\ref{fig:relaxtheory1}.

For $R^*/a\ll 1$, the four-atom loss mechanism is also suppressed. 
This suppression has the same origin (Pauli principle) as in the three-atom case: Four atoms consisting of two pairs of identical fermions have to approach each other to the recombination region. 
In this case we have $\eta_{\rm DD}\propto (R^*/a)^{2\nu_\text{4-body}+4}$. 
Here the power $\nu_\text{4-body}$ characterizes the scaling of the four-atom wave function in the universal region and can be inferred from the energy of four trapped fermions at unitarity: 
Reference~\cite{vonStecher2008eas} gives $\nu_\text{4-body}\approx 0.0,\, 0.3,$ and $0.5$ for mass ratios of 1, 4, and 8, respectively. 
Our calculation for the LiK mass ratio is consistent with this sequence.

\subsection{Comparison with experimental data}
Here we compare our theoretical predictions for the collisional loss-rate coefficients to our measured values, which we already presented in Figs.~\ref{Figure5} and \ref{Figure6}.
In our theoretical model, the three bare rate constants $\beta_\mathrm{D}^{(0)}$, $\beta_\mathrm{LiD}^{(0)}$, and $\beta_\mathrm{KD}^{(0)}$ are free parameters, and they can in principle be determined by fitting to the experimental data. 
Alternatively, estimates can be obtained from a simple quantum Langevin model \cite{Gao2010umf,Julienne2011uuc,Quenemer2011uiu}. 
This model uses only the van der Waals range of the corresponding atom-dimer or dimer-dimer interaction potential and assumes total absorption (loss) at shorter distances.

\begin{figure}
\centering
\includegraphics[clip,width=0.98\columnwidth]{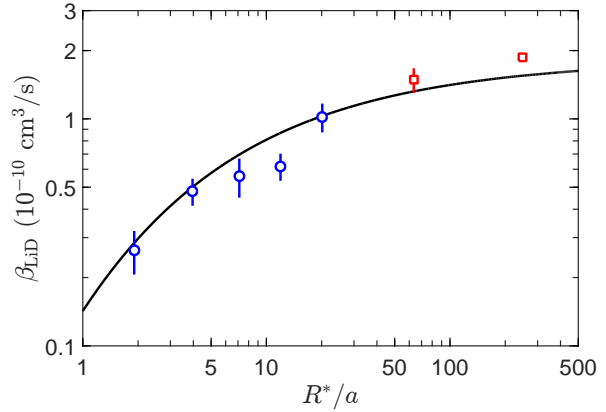}
\caption{(Color online) Atom-dimer loss-rate coefficient $\beta_\mathrm{LiD}$ as a function of $R^*/a$.
The experimental data (blue circles and red squares obtained with Li$|1\rangle$K$|3\rangle$ and Li$|1\rangle$K$|1\rangle$ dimers, respectively) are identical to the ones displayed in Fig.~\ref{Figure6}, with the unphysical negative value excluded.
The black solid line corresponds to a fit of our theoretical model to the data, yielding $\beta_\mathrm{LiD}^{(0)} =  1.8(2)\times10^{-10}\mathrm{cm}^{3}/\mathrm{s}$.}
\label{Figure8}
\end{figure}

For collisions of Li atoms with LiK dimers, the comparison is straightforward, since $\beta_\mathrm{LiD}^{(0)}$ is the only free parameter, which enters as a prefactor according to Eq.~(\ref{beta_AD}). 
Accordingly, we fit $\beta_\mathrm{LiD}(R^*/a) = \beta_\mathrm{LiD}^{(0)} \eta_\mathrm{LiD}(R^*/a)$ to the experimental data and extract the value $\beta_\mathrm{LiD}^{(0)} = 1.8(2)\times 10^{-10}\mathrm{cm}^3/\mathrm{s}$. 
The fit curve is shown as the black solid line in Fig.~\ref{Figure8} and shows that the theory matches the experimentally observed behavior very well. 
In particular, we can clearly confirm that the observed reduction of losses can be attributed to the Pauli suppression effect.

The value for $\beta_\mathrm{LiD}^{(0)}$ obtained from our fit analysis corresponds to about half of the value suggested by the quantum Langevin model amounting to $3.5\times 10^{-10}\mathrm{cm}^3/\mathrm{s}$.
Similar deviations have previously been observed in other experiments, in particular those involving light atoms \cite{Wang2013dfu}.

\begin{figure}
\centering
\includegraphics[clip,width=0.98\columnwidth]{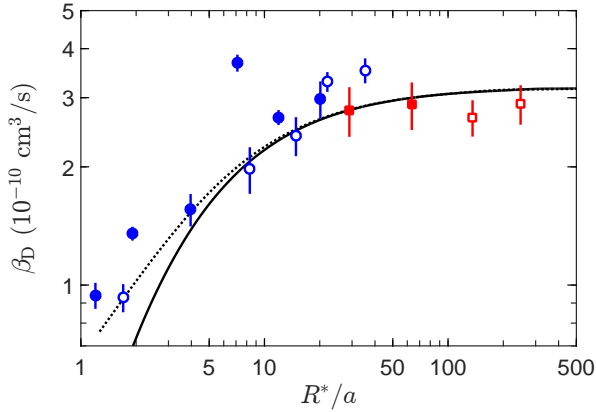}
\caption{(Color online) Total dimer-dimer loss-rate coefficient $\beta_\mathrm{D}$ as a function of $R^*/a$.
The experimental data (blue circles and red squares obtained with Li$|1\rangle$K$|3\rangle$ and Li$|1\rangle$K$|1\rangle$ dimers, respectively) are the same as those displayed in Fig.~\ref{Figure5}.
The solid line corresponds to a fit of our theoretical prediction for $\beta_\mathrm{D} = \beta_\mathrm{D}^{(0)} \eta_\mathrm{DD} + \beta_\mathrm{LiD}^{(0)} \tilde{\eta}_\mathrm{LiD}  + \beta_\mathrm{KD}^{(0)} \tilde{\eta}_\mathrm{KD}$ to the data, with $\beta_\mathrm{D}^{(0)}$ being the only free parameter (see text).
The dotted line corresponds to our prediction from an extended, finite-temperature theory \cite{UnPubfiniteT}.}
\label{Figure9}
\end{figure}

For collisions between dimers, the situation is more involved because of the three different channels -- Li-cc molecule, K-cc molecule, and cc molecule-cc molecule  
-- with the corresponding three free parameters $\beta_\mathrm{LiD}^{(0)}$, $\beta_\mathrm{KD}^{(0)}$, and $\beta_\mathrm{D}^{(0)}$, see Eq.~(\ref{beta_DD}).
According to our model, the dominant loss contribution is expected from the four-body channel. 
In order to extract the corresponding bare rate coefficient $\beta_\mathrm{D}^{(0)}$, we perform a one-parameter fit after fixing $\beta_\mathrm{LiD}^{(0)}$ to the measured value discussed before and fixing $\beta_\mathrm{KD}^{(0)}$ to the value $1.4\times 10^{-10}\mathrm{cm}^3/\mathrm{s}$ calculated within the quantum Langevin model \cite{Note3}. 
We finally obtain $\beta_\mathrm{D}^{(0)} = 3.2(6)\times 10^{-10}\mathrm{cm}^3/\mathrm{s}$, which we find to be very close the quantum-Langevin value, $\beta_\mathrm{D}^{(0)} = 3.0\times 10^{-10}\mathrm{cm}^3/\mathrm{s}$.
The resulting total decay rate $\beta_\mathrm{D}(R^*/a)$ is shown as the black solid line in Fig.~\ref{Figure9}.
Our theoretical approach reproduces the observed suppression of collisional relaxation as we approach the Feshbach resonance for $R^*/a \gtrsim 3$.

Closer to the Feshbach resonance ($R^*/a < 3$), we see clear deviations.
We ascribe this discrepancy to temperature effects, which become more prominent when $k_\mathrm{B} T$ is comparable to or larger than the dimer binding energy~\cite{Levinsen2011ada}. 
In this case, the identical fermions may more easily approach each other, thereby reducing the Pauli suppression factor.
A theoretical prediction obtained from a finite-temperature calculation for $T = 550\,$nK \cite{UnPubfiniteT} is shown as the black dotted line.
Including finite temperature into our theoretical approach improves the match of theory and experiment near the resonance.
From our finite-temperature calculations, we also find that corresponding effects on the collisions of Li atoms with LiK dimers, as discussed before, remain much smaller \cite{UnPubfiniteT}.

The good agreement between our theoretical approach and our experimentally obtained values for the loss coefficients validates the assumptions of our theoretical approach to collisional losses developed in Ref.~\cite{Levinsen2011ada}.
Furthermore, our results demonstrate that the bare rate coefficients can be well estimated by the value obtained from the quantum Langevin model.
The agreement with our measurements therefore suggests a predictive power of our theory applied to other Fermi-Fermi systems.

\section{Other potential Fermi-Fermi systems}
Fermi-Fermi systems that feature mass imbalance, collisional stability and tunable interactions may be created with mixtures other than $^{6}$Li-$^{40}$K. 
To date, Fermi degeneracy has been demonstrated for isotopes of eight chemical elements,
He \cite{Mcnamara2006dbf}, 
Li  \cite{Truscott2001oof, Schreck2001qbe}, 
K \cite{DeMarco1999oof},  
Cr \cite{Naylor2015cdf},  
Sr \cite{DeSalvo2010dfg, Tey2010ddb},  
Dy \cite{Lu2012qdd},  
Er \cite{Aikawa2014rfd},  
and
Yb \cite{Fukuhara2007dfg},
providing a wealth of possible combinations. 
We focus our attention to mixtures of $^{161}$Dy and $^{40}$K (mass ratio 4.0) and $^{53}$Cr and $^{6}$Li (8.8), and we discuss the corresponding suppression functions for collisional losses. 
Larger mass ratios (comparable or larger than 13.6) require an analysis beyond the scope of our present work. 
In this case, the Efimov \cite{Efimov1973elo} and other few-body effects \cite{Kartavtsev2007let, Castin2010fbe, Blume2012ufb} can lead to the appearance of new loss-rate features \cite{Marcelis2008cpo}.

\begin{figure*}
\centering
\includegraphics[clip,width=0.7408\columnwidth]{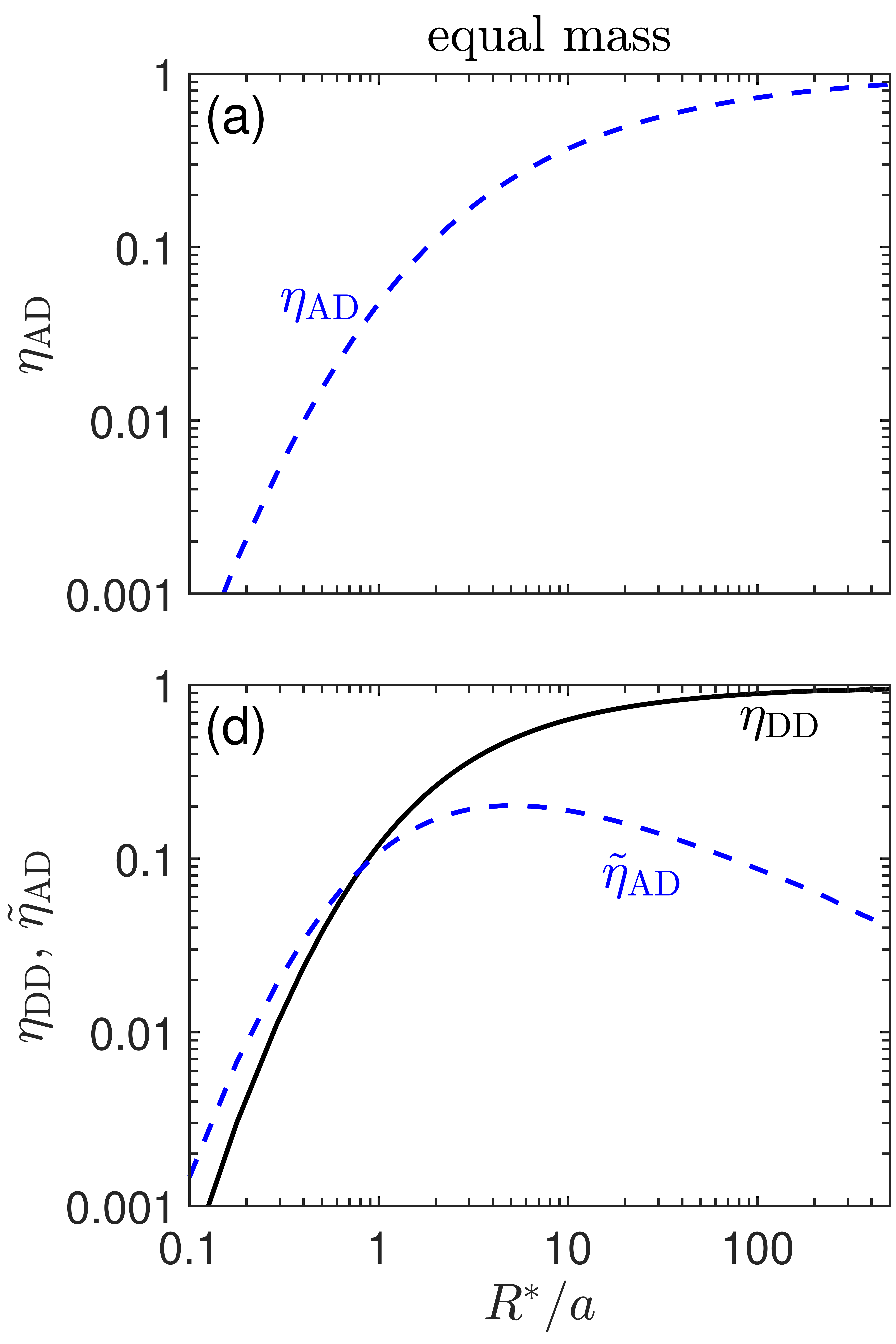}
\includegraphics[clip,width=0.6296\columnwidth]{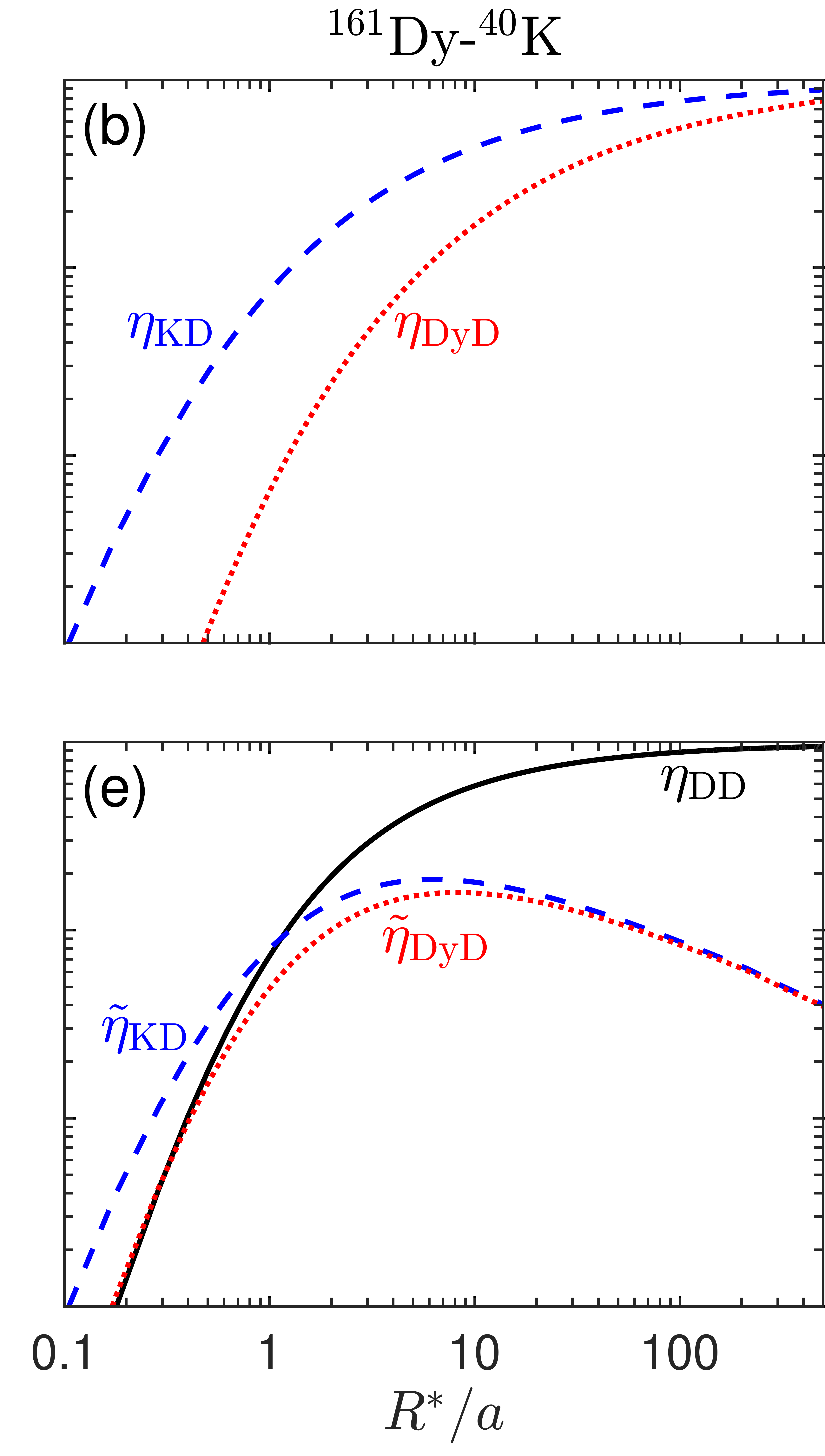}
\includegraphics[clip,width=0.6296\columnwidth]{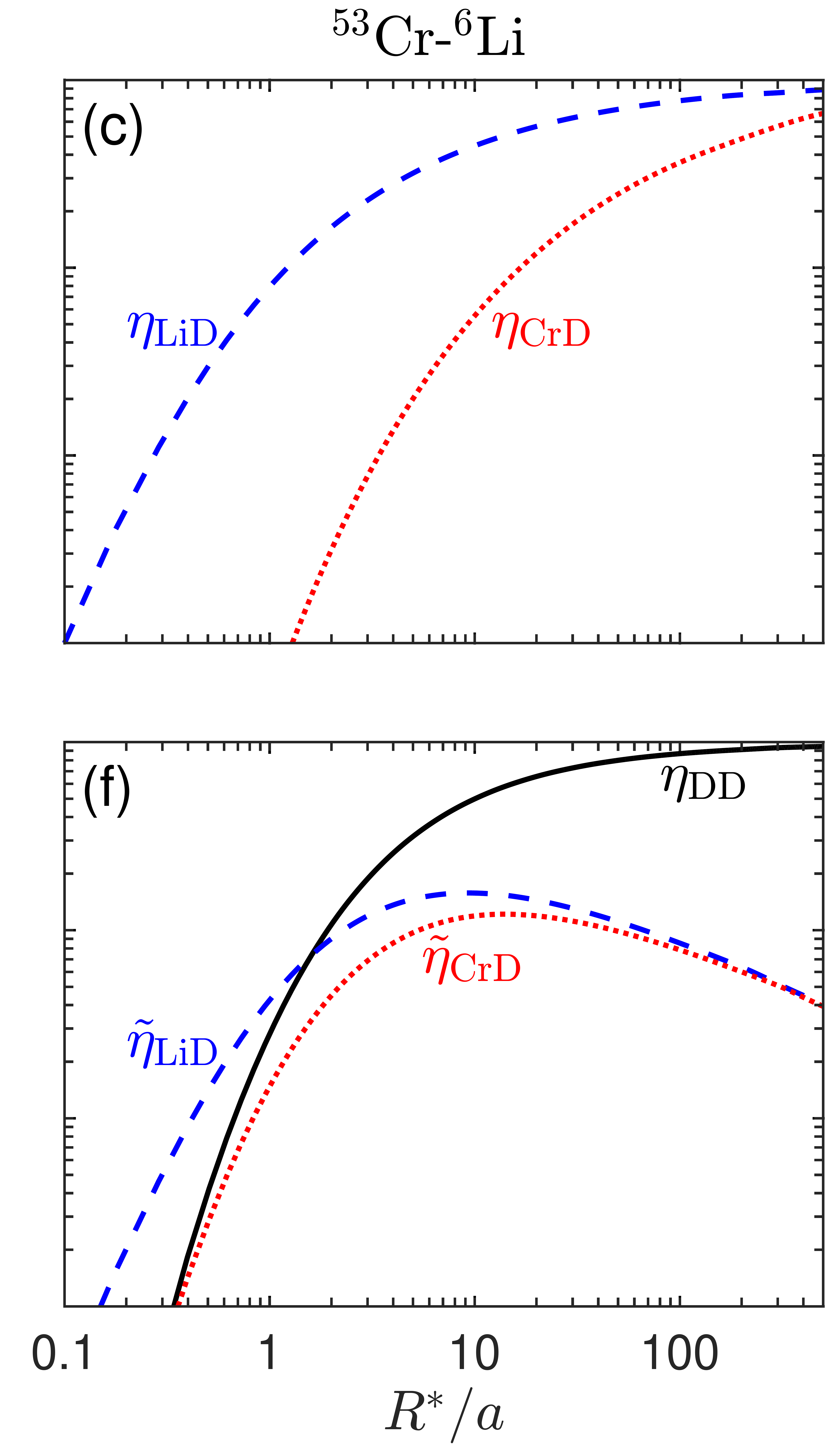}
\caption{(Color online) (a) Suppression functions for losses in (a,b,c) atom-dimer and (d,e,f) dimer-dimer collisions. 
We consider these for equal masses, Dy-K mixtures with a mass ratio of 4.0, and mixtures of Cr and Li where the mass ratio is 8.8.
}
\label{fig:relaxtheory}
\end{figure*}

The suppression functions for losses in atom-dimer collisions are shown in the upper panels of Fig.~\ref{fig:relaxtheory} for (a) the mass-balanced system, (b) the $^{161}$Dy-$^{40}$K mixture, and (c) the $^{53}$Cr-$^{6}$Li mixture.
We observe that mixtures of heavy-species atoms and dimers (red dotted lines) show a much stronger suppression compared to the mass-balanced case (blue dashed line in Fig.~\ref{fig:relaxtheory}(a)), which strengthens with increasing mass imbalance.
For the case of the Dy-K (Cr-Li) mixture, this increase amounts to almost one (two) orders of magnitude at $R^*/a$ corresponding to about 1.
On the contrary, the mixtures composed of light-species atoms and dimers (blue dashed lines in Fig.~\ref{fig:relaxtheory}(b) and (c)) show only a weak enhancement of losses as compared to the mass-balanced case, amounting to a factor of about $1.5$ for both the Dy-K and the Cr-Li mixture at $R^*/a = 1$.

The suppression functions for losses in collisions between dimers are shown in Fig.~\ref{fig:relaxtheory} (d), (e), and (f) for the equal-mass system, and the systems with a mass imbalance of 4 and 8.8, respectively.
All three contributions, from the light atom-dimer, from the heavy atom-dimer, and from the dimer-dimer part, shown in Figs.~\ref{fig:relaxtheory}(e) and (f) as the blue dashed, red dotted, and black solid lines, respectively, are significantly smaller than their equal-mass counterparts (blue dashed and black solid lines in Fig.~\ref{fig:relaxtheory}(d)).

In view of future experiments on strongly interacting Fermi-Fermi systems, we can now provide estimates for the minimum strength of a suitable Feshbach resonance. 
The conditions of the successful experiments with spin mixtures of $^{40}$K or $^{6}$Li suggest a minimum required suppression of losses by two orders of magnitudes for all possible channels. According to our theoretical results (Figs.~\ref{fig:relaxtheory1} and \ref{fig:relaxtheory}), this would correspond to a condition of $R^*/a \lesssim 0.3$ for all  mass-imbalanced mixtures considered. 
For the relevant scattering length we may take $a \approx 3000\,a_0$ as a typical value for dimers entering the strongly interacting Fermi gas regime. 
We thus obtain the approximate condition $R^* \lesssim 1000\,a_0$ for a Feshbach resonance to provide sufficient collisional stability. 

\section{Summary and Conclusion}
In a joint experimental and theoretical effort, we have investigated the stability of weakly bound dimers formed near narrow interspecies Feshbach resonances in Fermi-Fermi mixtures. 
In our laboratory system -- the mixture of $^6$Li and $^{40}$K atoms -- we have characterized the dependence of three different decay processes on the magnetic detuning from the Feshbach-resonance center. 
In dilute samples, spontaneous dissociation (one-body process) is observed for dimers composed of atoms that are not in the lowest spin channel, and the measured lifetimes are found to be in a full agreement with a previous theoretical prediction. 
In dense samples, we have measured the rate coefficients for inelastic dimer-dimer collisions as well as collisions of the lighter atomic species with the dimers. 
For all decay processes, we find a significant suppression when the resonance center is approached.

Our theoretical framework for the description of collisional losses near narrow Feshbach resonances is based on a model that has been developed in Ref.~\cite{Levinsen2011ada}. 
The basic idea is a separation of the problem into a long-range description of the three- and four-body kinematics and a simple relaxation model at short range. 
The reduction of collisional decay near the resonance center is described by corresponding suppression functions. 
In extension of previous work \cite{Levinsen2011ada}, we have calculated the suppression functions for all relevant loss channels in atom-dimer and dimer-dimer collisions.
The comparison of theoretical and experimental results for the mixture of $^6$Li and $^{40}$K shows excellent agreement, thus validating the assumptions of our theoretical model. 

The observed collisional suppression does not exceed a factor of about five, and thus stays far below what has been observed in homonuclear systems near broad resonances. 
Nevertheless, our present work shows that the $^6$Li-$^{40}$K system, in spite of the narrow nature of interspecies resonances \cite{Wille2008eau, Naik2011fri, Tiecke2010bfr}, can potentially exhibit a strong Pauli suppression of collisional losses, provided the density and resonance detunings can be substantially reduced. 
Under such conditions, spontaneous dissociation can be expected to become the dominant loss mechanism, with a strong effect on the system. 
This loss process could be avoided by choosing resonances in the lowest spin channel, which are all very narrow. 
The level of control required to manipulate the Li-K mixture at very low densities near the narrow resonances is very challenging, going far beyond typical conditions of the present Fermi gas experiments.

Other Fermi-Fermi mixtures are very promising for new experiments in the near future, and we have discussed the $^{161}$Dy-$^{40}$K case (mass ratio 4.0) and the $^{53}$Cr-$^{6}$Li (8.8) case as two illustrative examples. 
Efforts to realize these systems are under way in different laboratories, and their yet unknown interaction properties need to be explored. 
The suppression functions that we have calculated for the corresponding mass ratios provide a guide for identifying suitable Feshbach resonances in future experimental work. 
In general, our results suggest that broad Feshbach resonances are not necessarily required to obtain sufficient collisional stability. 
Instead, moderately narrow resonances are also promising for realizing new experimental model systems and for exploring the multifaceted many-body physics of fermionic mixtures \cite{Liu2003igs, Caldas2005pti, Iskin2006tsf, Lin2006ssf, Parish2007pfc, Iskin2008tif, Baranov2008spb, Bausmerth2009ccl, Gezerlis2009hlf, Baarsma2010pam, Braun2014pos, Braun2015zte}.

\begin{acknowledgments}
We thank J. Walraven and M. Zaccanti for stimulating discussions. 
The experimental team from Innsbruck acknowledges support by the Austrian Science Fund FWF within the Spezialforschungsbereich (SFB) FoQuS, project part P04 (F4004-N23). 
The research leading to the theoretical results received funding from the European Research Council (FR7/2007-2013 Grant Agreement No. 341197).
D.S.P. thanks IFRAF for support.
\end{acknowledgments}

\appendix*
\section{Theoretical approach to collisional decay}

\def\q{{\bf q}}
\def\p{{\bf p}}
\def\k{{\bf k}}
\def\u{{\bf u}}
\def\v{{\bf v}}
\def\0{{\bf 0}}
\def\Q{{\bf Q}}

\def\down{\downarrow}
\def\up{\uparrow}
\def\nn{\nonumber}

Here, we present our calculation of the probability to find an atom close to a closed-channel molecule in dimer-dimer collisions. 
As discussed in the main text, this probability allows us to extract the contribution from the corresponding relaxation channel to the dimer-dimer relaxation rate constant $\beta_\text{D}$.  
This extends our previous calculation of the cc molecule--cc molecule relaxation channel, as presented in Ref.~\cite{Levinsen2011ada}. 
Throughout this appendix we set $\hbar=1$ and work in a unit volume.

We now briefly recapitulate the theoretical description of our system. 
We consider two species of fermions labeled by $\sigma=\up,\down$ and employ the two-channel Hamiltonian~\cite{Timmermans1999fri}
\begin{align}
\hat H & = \sum_{\k,\sigma=\up,\down}\epsilon_{\k,\sigma}\hat 
a^\dagger_{\k,\sigma}\hat 
a_{\k,\sigma}+\sum_{\p}\left(\omega_0+\epsilon_{\p,\text{M}}\right)
\hat b^\dagger_\p \hat b_\p
\nn \\ & \hspace{-3mm}
+g\sum_{\k,\p}
\left(\hat b^\dagger_\p\hat a_{\frac{\p}2+\k,\up}\hat a_{\frac\p2-\k,\down}+
\hat b_\p\hat a^\dagger_{\frac\p2-\k,\down}\hat a^\dagger_{\frac\p2+\k,\up}\right).
\label{eq:hamiltonian}
\end{align}
Here, $\hat a^\dagger_{\k,\sigma}$ $(\hat a_{\k,\sigma})$ creates (annihilates) a spin $\sigma$ atom of mass $m_\sigma$ with momentum $\k$ and single-particle energy $\epsilon_{\k,\sigma}=\frac{k^2}{2m_\sigma}$. 
Likewise, $\hat b^\dagger_{\k}$ and $\hat b_{\k}$ are the creation and annihilation operators of the cc molecule with mass $M=m_\up+m_\down$, kinetic energy $\epsilon_{\k,\text{M}}=\frac{k^2}{2M}$, and detuning  $\omega_0$ from the $\up\down$ scattering threshold. 
The interaction between the atoms is mediated by the cc molecule as described by the last term of the Hamiltonian, where the strength $g$ of the interconversion term is taken to be constant up to a momentum cutoff $\Lambda$. 
The bare parameters of the model are related to the physical scales, the scattering length and length parameter $R^*$, through (see, e.g.,~\cite{Gurarie2007rpf})
\begin{align}
a=\frac{ m_r  g^2}{2\pi}\frac1{\frac{g^2 m_r \Lambda}{\pi^2}-\omega_0}, 
\hspace{1cm} R^*=\frac{\pi}{ m_r ^2g^2},
\label{eq:parameters}
\end{align}
with $ m_r =m_\up m_\down/M$ the reduced mass. The propagator of the atoms takes the form
\begin{align}
G_{\sigma}(\p,p_0) = \frac1{p_0-\epsilon_{\p,\sigma}+i0}, 
\end{align}
where the notation $+i0$ specifies that the pole of $p_0$ is shifted slightly into the lower half of the complex plane. 
The propagator of dimers is obtained by dressing the cc molecule propagator by pairs of free $\up\down$ atoms, resulting in
\begin{align}
D(\p,p_0) & \nn \\ & \hspace{-13mm}
= \frac{2\pi/ m_r }{2 m_r  R^*\!\left(p_0-\frac{p^2}{2M}+i0 \right)\!+\frac1{a}-\sqrt{2 m_r }
\sqrt{-p_0+\frac{p^2}{2M}\!-\!i0}}.
\label{eq:DressedMoleculePropagator}
\end{align}
At zero momentum, this has a pole at the dimer binding energy $\epsilon_0 = -(\sqrt{1+4R^*/a}-1)^2/(8 m_r  {R^*}^2)$.

\begin{figure*}
\centering
\includegraphics[clip,width=1.8\columnwidth]{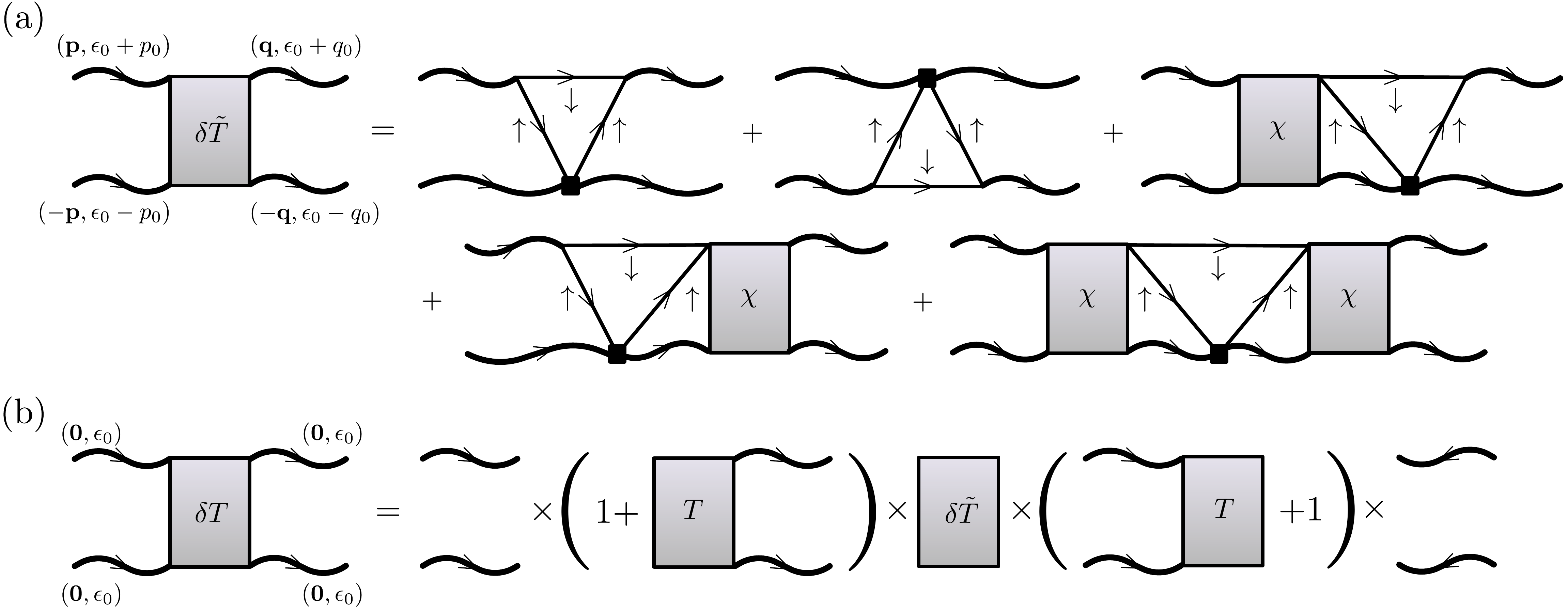}
\caption{Diagrams contributing to the $\up$ atom-cc molecule loss process in dimer-dimer collisions. 
The loss vertex \eqref{eq:Hprime} is depicted as a black square. Straight lines are atom propagators, while the dimer propagators are illustrated with wavy lines. 
All filled boxes represent sums of diagrams. 
(a) The two-dimer irreducible diagrams denoted $\delta \tilde T$ contributing to $\delta T$ can be obtained from the vertex $\chi$ (see text). 
(b) All diagrams in $\delta T$ can be obtained from the two-dimer irreducible diagrams by the use of the full dimer-dimer T matrix (see text). 
In both subfigures, the external dimer lines are for illustration only.}
\label{fig:diagrams}
\end{figure*}

To extract the relaxation rate for the three-atom process in a dimer-dimer collision, we introduce a weak short-range interaction potential between the $\up$ atom and the bare molecule in the Hamiltonian (see Ref.~\cite{Levinsen2011ada}):
\begin{align}
\delta \hat H_{\up\text{D}}=-i\Delta_{\up\text{D}}\sum_{{\bf Q},\k,\p}\hat b^\dag_{\p}\hat a^\dag_{\up,{\bf Q}-\p}\hat b_\k \hat a_{\up,{\bf Q}-\k}.
\label{eq:Hprime}
\end{align}
The coefficient $\Delta_{\up\text{D}}$ is related to the relaxation coupling constant through $g_{\up\text{D}}=-i \Delta_{\up\text{D}}$.
The probability to find the $\up$ atom close to the cc molecule in dimer-dimer scattering at zero collisional energy is then
\begin{align}
\tilde \eta_{\up\text{D}}(R^*/a)=-\text{Im}[\delta T(0)]/(2\Delta_{\up\text{D}}),
\end{align}
where $\delta T(0)$ is the change in the $s$-wave dimer-dimer scattering T matrix to linear order in $\Delta_{\up\text{D}}$. 
The change in the dimer-dimer scattering length to the same order is in turn $\delta a_\text{DD}=\delta T(0)M/(4\pi)$. 
We calculate this change diagrammatically as illustrated in Fig.~\ref{fig:diagrams}: 
First, we consider all diagrams contributing to $\delta T$ which are two-dimer irreducible (i.e.~do not have two dimers propagating simultaneously). 
We then include all two-dimer reducible processes by replacing the incoming and/or outgoing dimers by the full dimer-dimer T matrix.

Consider first the sum of diagrams in Fig.~\ref{fig:diagrams}(a) constituting all two-dimer irreducible contributions to $\delta T(0)$. 
Taking the incoming [outgoing] dimers to have four-momenta $(\pm\p,p_0+\epsilon_0)$ [$(\pm\q,q_0+\epsilon_0)$], we denote this sum by $\delta \tilde  T(p,p_0;q,q_0)$. 
This does not depend on the angle between $\p$ and $\q$ as we take the $s$-wave projection. 
Integrating over frequencies in the closed loops of the diagrams in Fig.~\ref{fig:diagrams}(a) yields for the two-dimer irreducible contribution to $\delta T(0)$:
\begin{widetext}
\begin{align}
\delta
&  \tilde T(p,p_0;q,q_0)= -2i\Delta_{\up\text{D}}g^2Z^2
\int\frac{d\Omega_\p}{4\pi}\int\frac{d\Omega_\q}{4\pi}
& \nn \\
& \hspace{-3mm} \times\left[
2\sum_\Q G_\up(\p-\Q,\epsilon_0+p_0-\epsilon_{\Q,\down})
  G_\up(\q-\Q,\epsilon_0+q_0-\epsilon_{\Q,\down})\right. \nn \\
& \, + \sum_{\p_1,\p_2}\chi(p,p_0;\p_1,\p_2)
D(\p_1+\p_2,2\epsilon_0-\epsilon_{\p_1,\up}-\epsilon_{\p_2,\down})
G_\up(\q-\p_1,\epsilon_0 +q_0-\epsilon_{\p_1,\up})
 \nn \\
& \, + \sum_{\p_1,\p_2}\chi(q,q_0;\p_1,\p_2)
D(\p_1+\p_2,2\epsilon_0-\epsilon_{\p_1,\up}-\epsilon_{\p_2,\down})
G_\up(\p-\p_1,\epsilon_0 +p_0-\epsilon_{\p_1,\up})
\nn \\
& \, \left. +
\sum_{\p_1,\p_2,\p'_2}
\chi(p,p_0;\p_1,\p_2)\chi(q,q_0;\q_1,\q_2)
D(\p_1+\p_2,2\epsilon_0-\epsilon_{\p_1,\up}-\epsilon_{\p_2,\down})
D(\p_1+\p_2',2\epsilon_0-\epsilon_{\p_1,\up}-\epsilon_{\p_2',\down})
\right],
\end{align}
where we integrate over the angles of $\p$ and $\q$. $Z=1-1/\sqrt{1+4R^*/a}$ is the dimer residue at the energy pole. 
The function $\chi(p,p_0;\p_1,\p_2)$ is the sum of all diagrams with two incoming dimers at four-momenta $(\pm \p,p_0+\epsilon_0)$, an outgoing $\up$ [$\down$] atom with $(\p_1,\epsilon_{\p_1,\up})$ $[(\p_2,\epsilon_{\p_2,\down})]$, and an outgoing dimer with $(-\p_1-\p_2,2\epsilon_0-\epsilon_{\p_1,\up}-\epsilon_{\p_2,\down})$. 
The sum is averaged over the angle of $\p$. $\chi$ satisfies an integral equation derived in Ref.~\cite{Levinsen2011ada}; 
for the expression we refer the reader to Eq.~(29) of that paper.

Finally, we relate $\delta T$ to the two-dimer irreducible diagrams by allowing for any number of dimer-dimer scattering events on the left and/or right side of $\delta\tilde T$, see Fig.~\ref{fig:diagrams}(b). 
The relation is
\begin{align}
\delta T(0)=&\delta \tilde T(0,0;0,0)+2\int \frac{i \,dp_0}{2\pi}\sum_\p F(p,p_0)
\delta \tilde T(p,p_0;0,0) +
\int \frac{i \,dp_0}{2\pi}\frac{i \,dq_0}{2\pi}
\sum_{\p,\q}
F(p,p_0) \delta \tilde T(p,p_0;q,q_0)
F(q,q_0),
\end{align}
where $F(p,p_0)\equiv \frac1{g^2Z}T(p,p_0) D(\p,p_0+\epsilon_0) D(-\p,-p_0+\epsilon_0)$ with $T(p,p_0)$ the dimer-dimer T matrix in the absence of the perturbation \eqref{eq:Hprime}. 
To avoid poles and branch cuts, the $p_0$ and $q_0$ integration contours are rotated to the imaginary axis. 
The dimer-dimer T matrix satisfies an integral equation derived in Ref.~\cite{Levinsen2011ada}, see Eq.~(28) of that paper.

\end{widetext}

\bibliographystyle{apsrev}

\end{document}